\documentclass[%
 aip,
 amsmath,amssymb,
 reprint,%
]{revtex4-2}

\usepackage{graphicx}
\usepackage{dcolumn}
\usepackage{bm}

\usepackage[utf8]{inputenc}
\usepackage[T1]{fontenc}
\usepackage{mathptmx}

\usepackage {xcolor}

\newcommand{\eps}{\varepsilon}

\newcommand{\om}{\omega}

\newcommand{\prt}{\partial}

\newcommand{\sn}{\mathrm{sn}}

\graphicspath{{Figures/}}

\begin{document}


\title{Motion of dark solitons in a non-uniform flow of Bose-Einstein condensate}

\author{S. K. Ivanov}
\author{A. M. Kamchatnov}

\affiliation{Institute of Spectroscopy,
Russian Academy of Sciences, Troitsk, Moscow, 108840, Russia}

\email{kamch@isan.troitsk.ru}

\date{\today}

\begin{abstract}
We study motion of dark solitons in a non-uniform one-dimensional flow of Bose-Einstein condensate.
Our approach is based on Hamiltonian mechanics applied to the particle-like behavior of dark
solitons in a slightly non-uniform and slowly changing surrounding. In one-dimensional geometry,
the condensate's wave function undergoes the jump-like behavior across the soliton and this leads to
generation of the counterflow in the background condensate. For correct description of
soliton's dynamics, the contributions of this counterflow to the momentum and energy of the
soliton are taken into account. The resulting Hamilton equations are reduced to the Newton-like
equation for the soliton's path and this Newton equation is solved in several typical situations.
The analytical results are confirmed by numerical calculations.
\end{abstract}



\maketitle

\begin{quotation}
The idea that localized solitary waves behave in external fields like point particles
of the classical mechanics is very old and it was studied and used in a number of
articles. However, its application to dynamics of dark solitons has some peculiarities
which are not still fully understood.  For example, from naive point of view a dark
soliton in a cloud of Bose-Einstein condensate (BEC) is just a dip in the density distribution
and, consequently, its motion must result in the motion of the center of mass in the
opposite direction. In reality both experiments and numerical solutions of the
Gross-Pitaevskii equation demonstrate different laws of motion of the dark soliton and
the center of mass of the cloud of a quasi-one-dimensional condensate confined in a
harmonic trap. This means that the motion of dark solitons is accompanied by some
counterflow in the BEC cloud which contributes to the location of its center of mass.
Correspondingly, the contributions of the counterflow to the momentum and the energy of
dark solitons must be taken into account. Previously, this idea was applied to
description of dark solitons motion in a quiescent condensate's cloud. In this paper,
we extend it to the general case of slightly non-uniform and slowly changing background
distributions. The resulting Hamilton equations for the soliton's motion are reduced to
the Newton-like equation and this equation is solved for several typical situations of
the background evolutions, such as a rarefaction wave, a self-similar expansion of the
BEC cloud, a hydraulic flow of BEC past a wide obstacle. Solutions of the Newton equation
agree very well with numerical solutions of the full Gross-Pitaevskii equation.
\end{quotation}


\section{Introduction}\label{intro}

The name of {\it solitons} was coined by Zabusky and Kruskal \cite{zk-08}  due to their
particle-like behavior: two solitons interact elastically and pass through one another
without losing their identity. This concept of solitons as particle-like nonlinear wave
excitations was confirmed and amplified by the properties of solitons propagation in a
non-uniform external field (see, e.g., \cite{km-77,km-78,kn-78,kosevich-90}). In particular,
bright soliton clouds of attractive BEC oscillate in the external
trap potential $U=m\omega_0^2x^2/2$ ($m$ is an atomic mass) with the trap frequency
$\omega_0$. However, behavior of dark solitons in repulsive BEC has some peculiarities.
For example, in case of BEC confined in a quasi-one-dimensional (1D) harmonic
potential trap, they propagate along a stationary Thomas-Fermi profile of the density
and oscillate with the frequency $\om_0/\sqrt{2}$, although the center of mass of the
condensate still oscillates with the trap frequency $\om_0$ (see \cite{ba-2000}).
This means that the counterflow of the condensate should be taken into account. Such a
counterflow is caused by the condensate's phase jump across a dark soliton
\cite{shevchenko-88,ps-2003} and it leads to the difference between the canonical momentum $p$
of the soliton quasi-particle and its `naive' mechanical momentum \cite{ky-94,kk-95}.
As a result, the canonical momentum $p$ of a dark soliton, its energy $\eps$ and
its velocity $V$ are related by the classical formula
\begin{equation}\label{eq0}
  V=\frac{\prt\eps}{\prt p}
\end{equation}
of Hamiltonian mechanics, and this explicitly demonstrates that a dark soliton behaves
like an effective particle (quasi-particle). In particular, conservation of the energy
$\eps$ in a stationary field leads immediately \cite{kp-04} to the frequency
$\om_0/\sqrt{2}$ of dark solitons oscillations in a harmonic potential. Similar approach
yields easily the theory of evolution of ring solitons in cylindrically symmetrical traps
\cite{kk-10} and this theory agrees very well with numerical simulations.

The aim of this paper is to extend the above approach to situations with a non-uniform and
stationary or time-dependent flow of BEC. The Hamilton equations are reduced to
the Newton-like equation and in this way we reproduce the results obtained in Ref.~\cite{ba-2000}
by the perturbation method as well as the results derived in Ref.~\cite{she-18} from the
soliton limit of the Whitham modulation equations. In a stationary case, the Hamiltonian approach
provides the non-trivial energy conservation law.

We illustrate our approach by solutions of the dark soliton motion equation for such
typical situations as its motion along a rarefaction wave \cite{she-18}, along a self-similar
expanding cloud of BEC \cite{talanov-65,bkk-03}, along a stationary 1D flow of BEC past
an obstacle in the subcritical regime \cite{hakim-97,legk-09}. Solutions of the Newton equation
agree very well with numerical solutions of the Gross-Pitaevskii (GP) equation.


\section{Hamiltonian approach to dark solitons motion}\label{sec2}

One-dimensional dynamics of BEC
is described with high accuracy by the GP equation
which can be written in the non-dimensional form as
\begin{equation}\label{eq1}
	i\psi_t+\frac12\psi_{xx}-|\psi|^2\psi=U(x)\psi,
\end{equation}
where $U(x)$ is the external potential. Transition from the BEC wave
function $\psi$ to the more convenient hydrodynamic variables, namely, the condensate's
density $\rho$ and its flow velocity $u$, is performed by means of the substitution
\begin{equation}\label{eq2}
	\psi(x,t)=\sqrt{\rho(x,t)}\exp\left({i}\int^x u(x',t)dx'\right),
\end{equation}
so that the GP equation is cast to the system
\begin{equation}\label{eq3}
	\begin{split}
		&\rho_t+(\rho u)_x=0,\\
		&u_t+uu_x+\rho_x+\left[\frac{\rho_x^2}{8\rho^2}
		-\frac{ \rho_{xx}}{4\rho}\right]_x= -U_x.
	\end{split}
\end{equation}

We assume that the external potential $U(x)$ forms a non-uniform distribution of the
background density which changes at the characteristic distances of the order of magnitude
much greater then the healing length of BEC. Since the typical soliton's width has the order
of magnitude of the healing length, in derivation of the soliton solution with account of
the flow velocity we can neglect the potential term $U(x)$ in the above equations.
It is well known (see, e.g., \cite{kamch-2000}) that the traveling wave solution of
Eqs.~(\ref{eq3}) without the potential
term can be written in the form ($\rho=\rho(\xi)$, $u=u(\xi)$, $\xi=x-Vt$)
\begin{equation}
	\begin{split}
		\rho&=\rho_1+(\rho_2-\rho_1)\sn^2\left(\sqrt{\rho_3-\rho_1}(x-Vt),m\right),\\
		u &= V\pm\frac{\sqrt{\rho_1\rho_2\rho_3}}{\rho}, \quad \rho_1\le\rho_2\le\rho_3,
	\end{split}
\end{equation}
where
\begin{equation}\label{}
	m=\frac{\rho_2-\rho_1}{\rho_3-\rho_1}
\end{equation}
and $\rho(\xi)$ oscillates in the interval
\begin{equation}\label{}
	\quad \rho_1\le\rho\le\rho_2.
\end{equation}
Two signs in the formula for $u(\xi)$ correspond to two directions of the velocity of the
background flow along which the wave propagates. We have the same density profiles for both
directions of the flow.

We are interested in the soliton solution, so we turn to the soliton limit
$\rho_3\to\rho_2$ ($m\to1$) and obtain
\begin{equation}\label{eq9}
	\begin{split}
		\rho&=\rho_2-\frac{\rho_2-\rho_1}{\cosh^2\left[\sqrt{\rho_2-\rho_1}(x-Vt)\right]}, \\
		u&=V\pm\frac{\rho_2\sqrt{\rho_1}}{\rho}.
	\end{split}
\end{equation}
Far enough from the soliton's center the distributions tend to their asymptotic background values
$\rho\to\rho_0=\rho_2$, $u\to u_0$, that is $\rho_1=(V-u_0)^2$, so we get the soliton solution
in terms of the physical variables
\begin{equation}\label{eq92}
	\begin{split}
		\rho&=\rho_0-\frac{\rho_0-(V-u_0)^2}{\cosh^2\left[\sqrt{\rho_0-(V-u_0)^2}(x-Vt)\right]}, \\
		u&=V-\frac{\rho_0(V-u_0)}{\rho}.
	\end{split}
\end{equation}
Consequently, the absolute value of the soliton's speed relative to the BEC flow is smaller then
the sound velocity $c_0 = \sqrt{\rho_0}$ in a uniform condensate with the density $\rho_0$:
\begin{equation}\label{eq92a}
  |u_0-c_0|\leq V\leq u_0+c_0,
\end{equation}
where $V$ is defined in the `laboratory' reference system. As is clear from Eq.~(\ref{eq92}),
the amplitude of the dark soliton is related with its velocity by the formula
\begin{equation}\label{eq4-18}
  a=\rho_0-(V-u_0)^2.
\end{equation}

We consider a dark soliton as a quasi-particle \cite{kp-04}, that is a localized excitation
propagating through the moving condensate. To describe dynamics of such a soliton, we need to find
expressions for its energy and canonical momentum. As is known \cite{shevchenko-88},
the excitation of a dark soliton in BEC is accompanied  by generation of the counterflow in the
background condensate: this counterflow compensates the jump of the phase in BEC $\Delta\phi=\int udx$
corresponding to the solution (\ref{eq92}). Consequently, the canonical momentum of a dark
soliton must include the term resulting from the phase jump contribution. In case of the
background at rest the expression for the canonical momentum reads
\begin{equation}\label{t4-119.4_s}
  p=-2V\sqrt{\rho_0-V^2}+2\rho_0\arccos\frac{V}{\sqrt{\rho_0}}.
\end{equation}
If the background moves with velocity $u_0$, then we have to subtract $u_0$ from the soliton's
velocity $V$ to obtain the expression
\begin{equation}\label{t4-119.4}
\begin{split}
  p=&-2(V-u_0)\sqrt{\rho_0-(V-u_0)^2}\\
  &+2\rho_0\arccos\frac{V-u_0}{\sqrt{\rho_0}}
  \end{split}
\end{equation}
for the canonical momentum of dark solitons in the moving background. Then from Eq.~(\ref{eq0})
we have
$$
\eps=\int Vdp=\int V\frac{dp}{dV}dV
$$
and substitution of (\ref{t4-119.4}) followed by simple integration yields
\begin{equation}\label{sol-energy}
  \eps=\frac43\left[\rho_0-(V-u_0)^2\right]^{3/2}+u_0p\equiv \eps^{(0)}+u_0p,
\end{equation}
where $\eps^{(0)}(V)=\frac43(\rho_0-V^2)^{3/2}$ is the well-known expression for the
dark soliton's energy in the quiescent BEC. Formula (\ref{sol-energy}) corresponds to
the Galileo transformation of the quasi-particle's energy in agreement with the Landau
approach to the superfluidity theory \cite{ps-2003}. The last term in Eq.~(\ref{sol-energy})
can also be considered as the Doppler shift of the frequency due to motion of the medium.

So far we have considered dark solitons moving along a uniform condensate with the density $\rho_0$
and the flow velocity $u_0$. Now we assume that the distributions of the density $\rho=\rho(x,t)$
and the flow velocity $u=u(x,t)$ are slow functions of the space coordinate $x$ and time $t$,
that is they change little along the distance of the order of magnitude of the soliton's width.
Then we can introduce the coordinate $x=x(t)$ of the soliton and in the main approximation
$\rho(x(t),t)$, $u(x(t),t)$ represent the density and the flow velocity of the background
distributions at the soliton's location.
In this approximation, if we make the replacements
$$
\rho_0\mapsto\rho(x,t),\quad u_0\mapsto u(x,t),\quad V\mapsto\dot{x}(t)=\frac{dx}{dt}
$$
in Eqs.~(\ref{t4-119.4}) and (\ref{sol-energy}), then we arrive at the expressions for the
canonical momentum
\begin{equation}\label{h1}
  p=-2(\dot{x}-u)\sqrt{\rho-(\dot{x}-u)^2}+2\rho\arccos\frac{\dot{x}-u}{\sqrt{\rho}}
\end{equation}
and the energy
\begin{equation}\label{h2}
  \begin{split}
  \eps=&\frac43\left[\rho-(\dot{x}-u)^2\right]^{3/2}-2u(\dot{x}-u)\sqrt{\rho-(\dot{x}-u)^2}\\
 & +2u\rho\arccos\frac{\dot{x}-u}{\sqrt{\rho}}
  \end{split}
\end{equation}
of the particle-like dark soliton located at the moment of time $t$ at the point $x$,
where $\rho=\rho(x,t)$, $u=u(x,t)$. Hamilton equations for the dark soliton motion have the
standard form
\begin{equation}\label{h3}
  \frac{dx}{dt}=\left(\frac{\prt\eps}{\prt p}\right)_{x},\qquad
  \frac{dp}{dt}=-\left(\frac{\prt\eps}{\prt x}\right)_{p},
\end{equation}
where it is implied that the velocity $\dot{x}$ is excluded from the energy (\ref{h2})
with the use of Eq.~(\ref{h1}).

Since it is impossible to express the energy (\ref{h2}) as a function of $p$ and $x$ in an
explicit form, it is convenient to transform the Hamilton equations (\ref{h3}) to the
Newton-like equation for the soliton's path $x(t)$. To this end, we differentiate (\ref{h1})
with respect to time $t$ and find the expression for the left-hand side of the second
equation (\ref{h3}),
\begin{equation}\label{h4}
\begin{split}
  \frac{dp}{dt}=&-4(\ddot{x}-u_x\dot{x}-u_t)\sqrt{\rho-(\dot{x}-u)^2}\\
  &+ 2(\rho_x\dot{x}+\rho_t)\arccos\frac{\dot{x}-u}{\sqrt{\rho}}.
  \end{split}
\end{equation}
In Hamiltonian mechanics the velocity $\dot{x}=V$ is considered as a function $V=V(p,x)$
of the momentum $p$ and coordinate $x$ defined implicitly by Eq.~(\ref{h1}). Therefore the
derivative of the right-hand side of the second equation (\ref{h3}) can be written as
\begin{equation}\label{h5}
\begin{split}
  \left(\frac{\prt\eps}{\prt x}\right)_{p}=&\left.\frac{\prt\eps^{(0)}}{\prt\rho}\right|_{V,u}\rho_x
  + \left.\frac{\prt\eps^{(0)}}{\prt u}\right|_{V,\rho}u_x\\
  &+ \left.\frac{\prt\eps^{(0)}}{\prt V}\right|_{\rho,u}\cdot\left.\frac{\prt V}{\prt x}\right|_p+u_xp,
  \end{split}
\end{equation}
where $\eps^{(0)}=\frac43[\rho-(V-u)^2]^{3/2}$. The derivatives of $\eps^{(0)}$ with respect to
$\rho,u,V$ are calculated without any difficulty and the derivative $\left.\frac{\prt V}{\prt x}\right|_p$
is to be calculated by differentiation of Eq.~(\ref{h1}) with $\dot{x}=V(p,x)$ under the condition
that $p=\mathrm{const}$. After simple calculation we obtain
\begin{equation}\label{h6}
  \left(\frac{\prt V}{\prt x}\right)_{p}=u_x+\frac{\rho_x}{2\sqrt{\rho-(V-u)^2}}\arccos\frac{V-u}{\sqrt{\rho}}.
\end{equation}
Substitution of all derivatives into Eq.~(\ref{h5}) followed by substitution of Eqs.~(\ref{h4}) and
(\ref{h5}) into the second equation (\ref{h3}) yields the equation
\begin{equation}\label{t6-47.10}
\begin{split}
  &2\ddot{x}=\rho_x+(u+\dot{x})u_x+2u_t\\
  &+\frac{\rho_t+(\rho u)_x}{\sqrt{\rho-(V-u)^2}}\arccos\frac{V-u}{\sqrt{\rho}}.
  \end{split}
\end{equation}
This equation describes the dynamics of a soliton with account the source (pumping) and absorbtion of
the condensate. Since in our case we assume that a dark
soliton propagates along a smooth background whose evolution obeys the dispersionless limit
of Eqs.~(\ref{eq3}),
\begin{equation}\label{t6-47.11}
  \rho_t+(\rho u)_x=0,\qquad u_t+uu_x+\rho_x=-U_x,
\end{equation}
we can cast Eq.~(\ref{t6-47.10}) to the form
\begin{equation}\label{t6-47.12}
  2\ddot{x}=-2U_x-\rho_x+(\dot{x}-u)u_x
\end{equation}
or
\begin{equation}\label{t6-47.13}
  2\ddot{x}=-U_x+u_t+u_x\dot{x}=-U_x+\dot{u}.
\end{equation}
Equation (\ref{t6-47.13}) was derived in framework of the perturbation theory in Ref.~\cite{ba-2000}.

Let us list several important particular cases.

{\it (i)} If there is no external potential and a dark soliton propagates along a large scale wave
obeying the Euler equations (\ref{t6-47.11}) with $U=0$, then Eq.~(\ref{t6-47.13}) gives at once
the integral of motion
\begin{equation}\label{t6-47.14}
  2\dot{x}-u=\mathrm{const},
\end{equation}
which constant value is determined by the initial velocity $V_0$ and the value of the flow velocity
$u(x_0)$ at the initial point $x_0$ of the soliton's path. This integral follows also from the
soliton limit of the Whitham equations for the GP equation periodic solutions with slowly
changing parameters (see Ref.~\cite{she-18}).

{\it (ii)} If the flow is stationary, that is the distributions $\rho=\rho(x)$, $u=u(x)$ do not depend
on time $t$, then the motion equation
\begin{equation}\label{t6-47.12b}
  2\ddot{x}=-U_x+u_x\dot{x}
\end{equation}
has the integral of energy $\eps(\dot{x},x)=\mathrm{const}$, see Eq.~(\ref{h2}), as it
follows immediately from the Hamilton equations (\ref{h3}).

{\it (iii)} If there is no flow ($u=0$) and a dark soliton propagates along the stationary
Thomas-Fermi distribution $\rho(x)+U(x)=\mathrm{const}$, then its motion obeys the Newton
equation
\begin{equation}\label{h7}
  2\ddot{x}=-U_x.
\end{equation}
In case of condensate confined in a harmonic trap $U(x)=\omega_0^2x^2/2$ such a soliton
oscillates with the frequency $\omega_0/\sqrt{2}$ as was predicted in Ref.~\cite{ba-2000}
and confirmed in the experiments \cite{becker-08,weller-08}.

Now we turn to consideration of several typical examples a dark soliton propagation
along a non-uniform background.


\begin{figure*}[t]
	\begin{center}
		\includegraphics[width=\linewidth]{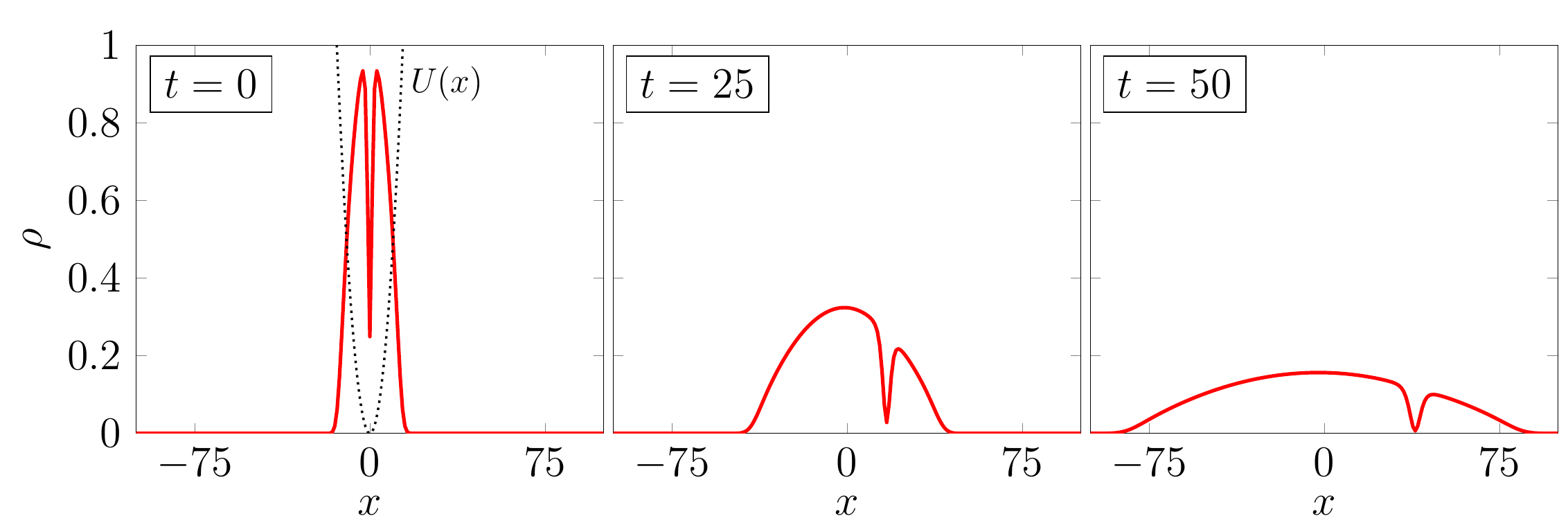}
		\caption{
			Dynamics of a dark soliton moving along expanding BEC. At $t=0$, the condensate is
trapped by a harmonic potential $U(x)=\omega_0^2x^2/2$ (dotted curve).
Here $\mu=1$, $\omega_0=0.1$ and $V_0=0.5$.
		}
		\label{fig1}
	\end{center}
\end{figure*}

\begin{figure}[t]
	\begin{center}
		\includegraphics[width=7cm]{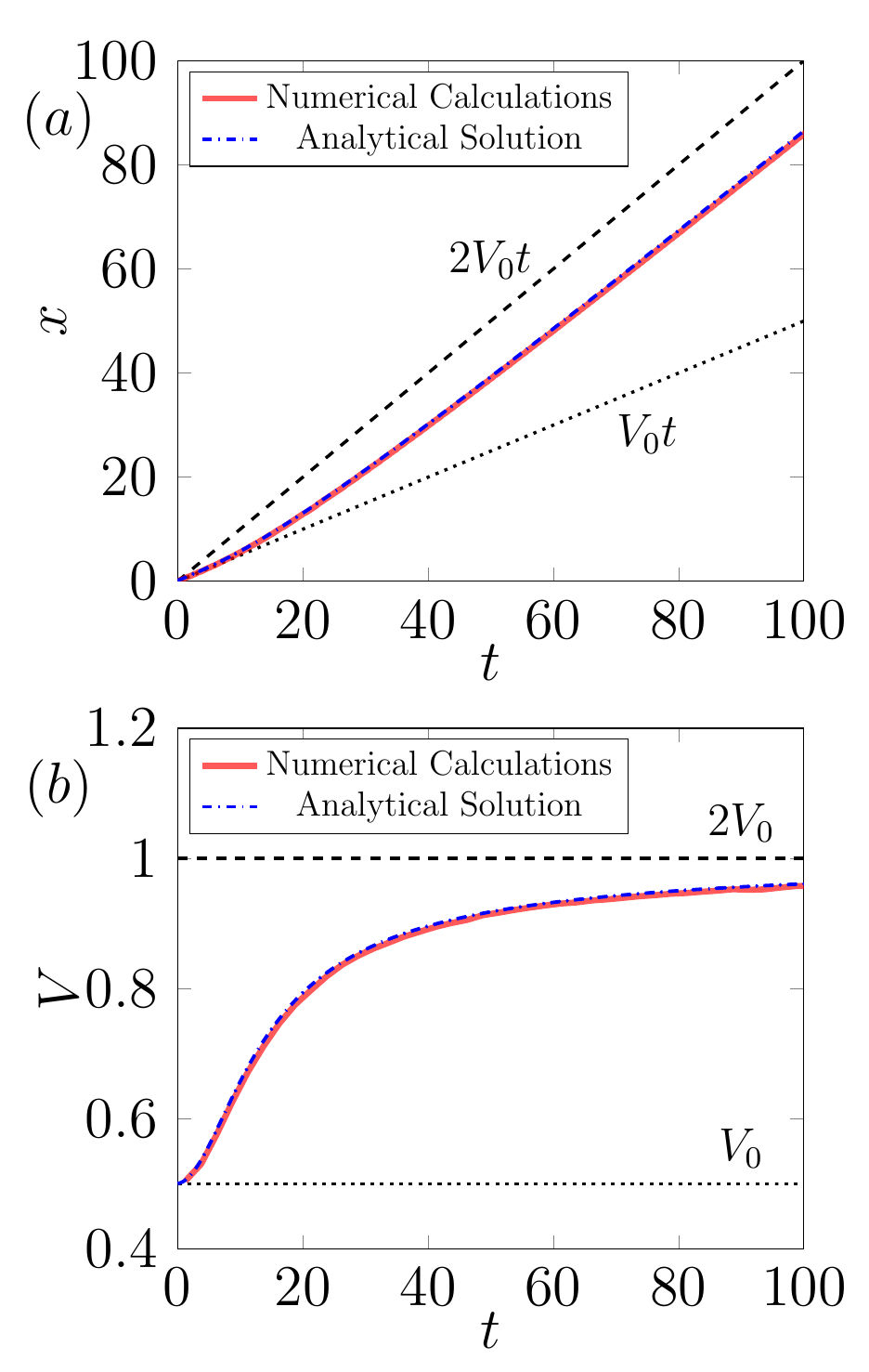}
		\caption{
			(a) Dark soliton's path $x(t)$ during its motion in the expanding condensate.
			(b) Velocity of the dark soliton $V=\dot{x}(t)$ as a function of time.
			The red curves show the results of numerical calculations and the dash-dotted blue curves
show the analytical solution. Dotted lines correspond to $V=V_0$ and dashed lines correspond
to $V=2V_0$. Here $\mu=1$, $\omega_0=0.1$ and $V_0=0.5$.
		}
		\label{fig2}
	\end{center}
\end{figure}


\section{Motion of dark solitons along a rarefaction wave}

One of the simplest non-uniform time-dependent solutions of equations (\ref{t6-47.11}) for the
background flow with $U=0$ is a rarefaction wave. Although the motion of dark solitons
along such a wave has already been considered in Ref.~\cite{she-18} by the Whitham method,
we shall consider it briefly for completeness by our method with some additions and
simplifications.

First, we introduce the Riemann invariants
\begin{equation}\label{h8}
  r_+=\frac{u}2+\sqrt{\rho},\qquad r_-=\frac{u}2-\sqrt{\rho},
\end{equation}
and transform Eqs.~(\ref{t6-47.11}) with $U=0$ to the diagonal Riemann form
\begin{equation}\label{h9}
  \frac{\prt r_-}{\prt t}+v_-\frac{\prt r_-}{\prt x}=0,\qquad
  \frac{\prt r_+}{\prt t}+v_+\frac{\prt r_+}{\prt x}=0,
\end{equation}
where
\begin{equation}\label{h10}
  v_-=\frac12(3r_-+r_+),\qquad v_+=\frac12(r_-+3r_+).
\end{equation}
The physical variables are related with the Riemann invariants by the formulas
\begin{equation}\label{h11}
  u=r_-+r_+,\qquad \rho=\frac14(t_+-r_-)^2.
\end{equation}

A rarefaction wave is a particular self-similar solution of Eqs.~(\ref{h9}) with one of
the Riemann invariants constant. To be definite, we assume that $r_-=\mathrm{const}$ and
the self-similar rarefaction wave evolves from the initial discontinuity, located at $x=0$,
of the Riemann invariant $r_+=u/2+\sqrt{\rho}$. Let the density equal to $\rho_-$ and
the flow velocity equal to zero, $u_-=0$ on the left side of the initial discontinuity.
We denote the density
on the right side of the discontinuity as $\rho_+$, $\rho_+>\rho_-$, and then from the condition
that $r_-$ is constant across the discontinuity, $r_-=u/2-\sqrt{\rho}=-\sqrt{\rho_-}$, we find
the flow velocity on the right from the discontinuity is equal to
\begin{equation}\label{t6-48.1}
  u_+=2(\sqrt{\rho_+}-\sqrt{\rho_-}).
\end{equation}
The first equation (\ref{h9}) is already satisfied and the self-similar solution $r_+=r_+(x/t)$
of the second equation gives $v_+=(3r_++r_-)/2=x/t$, or, with account of $r_-=-\sqrt{\rho_-}$,
we get $r_+=u+\sqrt{\rho_-}$. As a result we obtain the distribution of the flow velocity in
the rarefaction wave
\begin{equation}\label{t6-48.2}
  u(x,t)=\left\{
  \begin{array}{ll}
  0,\qquad & x<c_-t,\\
  \frac23\left(\frac{x}t-c_-\right),\qquad & c_-t<x<(3c_+-2c_-)t,\\
  2(c_+-c_-),\qquad & x>(3c_+-2c_-)t,
  \end{array}
  \right.
\end{equation}
where $c_{\pm}=\sqrt{\rho_{\pm}}$.

Let at the initial moment of time $t=0$ a dark soliton be located on the right from the discontinuity
at the point $x_+>0$, let it move with the velocity $V_+$, so that its amplitude equals to (see Eq.~(\ref{eq92}))
\begin{equation}\label{t6-48.3}
  a_+=\rho_+-(V_+-u_+)^2.
\end{equation}
Since the right edge of the rarefaction wave propagates with velocity $3c_+-2c_-$, it catches up
the soliton at the moment
\begin{equation}\label{t6-48.4}
  t_+=\frac{x_+}{3c_+-2c_--V_+}
\end{equation}
provided the denominator of this formula is positive what will be assumed in what follows.

For $t>t_+$ the soliton moves along the rarefaction wave and we can use the conservation law
(\ref{t6-47.14}), which in our case becomes the linear differential equation
\begin{equation}\label{t6-49.6}
  \frac{dx}{dt}-\frac{x}{3t}=V_+-c_++\frac23c_-.
\end{equation}
It can be readily solved with the initial condition  $x(t_+)=x_++V_+t_+$, so we get the formula
for the soliton's path,
\begin{equation}\label{t6-49.7}
  x(t)=\left(V_+-\frac{x_+}{2t_+}\right)t+\frac{3x_+}{2}\left(\frac{t}{t_+}\right)^{1/3},
  \qquad t>t_+.
\end{equation}
The soliton reaches the left edge of the rarefaction wave at the moment $t_-$ defined by the
condition $x(t_-)=c_-t_-$ which gives
\begin{equation}\label{t6-49.8}
  t_-=\frac1{\sqrt{t_+}}\left(\frac{x_+}{c_+-V_+}\right)^{3/2}.
\end{equation}
Consequently, the soliton passes through the rarefaction wave if only $c_+>V_+$ and in this case
Eq.~(\ref{t6-47.14}) gives at once that for $t>t_-$ the soliton's velocity becomes equal to
$V_-=c_--(c_+-V_+)$. Its amplitude $a_-=\rho_--V_-^2$ can be transformed with the help of
Eq.~(\ref{t6-48.3}) to
\begin{equation}\label{t6-49.9}
  a_-=a_+-2(c_+-c_-)\left(c_+\pm\sqrt{c_+^2-a_+}\right),
\end{equation}
where the choice of the sign is determined by the sign of the relative velocity $V_+-u_+$.

If $c_+=V_+$, then the soliton's path is given by
\begin{equation}\label{t6-49.10}
  x(t)=c_-t+\frac{3x_+}{2}\left(\frac{t}{t_+}\right)^{1/3},\quad t>t_+
  =\frac{x_+}{2(c_+-c_-)},
\end{equation}
that is it propagates ahead of the rear (left) edge of the rarefaction wave.

If $c_+<V_+$, the soliton also remains forever inside the rarefaction wave, but this time
its velocity tends according to Eq.~(\ref{t6-49.6}) to the constant value
\begin{equation}\label{t6-49.11}
  V_{\infty}=c_-+\frac32(V_+-c_+),
\end{equation}
which is greater that the velocity $c_-$ of the left edge of the rarefaction wave. Along the limiting
trajectory $x=V_{\infty}t$ the background flow has the values $u_{\infty}=V_+-c_+$,
$\rho_{\infty}=[c_-+(V_+-c_+)/2]^2$. Hence, the relative soliton's velocity
$V_{\infty}-u_{\infty}$ is equal to the local sound velocity $c_{\infty}=\sqrt{\rho_{\infty}}$,
that is the soliton's amplitude tends to zero in the limit $t\to\infty$.


\section{Dark soliton's motion along expanding condensate}

Now we assume that the initial background state is formed by a harmonic potential $U(x)=\omega_0^2x^2/2$ (in non-dimensional variables $m=1$) and the distribution of density is given by the Thomas-Fermi formula $\rho(x,0)=(2\mu-\omega_0^2x^2)/2, |x|\leq\sqrt{2\mu}/\omega_0$, where $\mu$ is the chemical potential of BEC, and the initial flow velocity is equal to zero everywhere.
At the initial moment of time $t=0$ the trap is switched off and the condensate starts its
expansion. The dispersionless equations (\ref{t6-47.11}) can be solved exactly in this case,
the solution was found in Ref.~\cite{talanov-65} for the focusing NLS equation and it was
adapted to the present repulsive BEC situation in Ref.~\cite{bkk-03}. It is given by the
formulas
\begin{equation}\label{h12}
\begin{split}
  \rho(x,t)&=\frac1{f(t)}\left(\mu-\frac{x^2}{\phi_0^2f^2(t)}\right),\qquad u(x,t)=xg(t)\\
  g(t)&=\frac{f'(t)}{f(t)}=\frac{2}{\phi_0}\sqrt{\frac{f(t)-1}{f^3(t)}}, \qquad  \phi_0=\frac{\sqrt{2}}{\omega_0},
\end{split}
\end{equation}
where the function $f(t)$ satisfies the differential equation
\begin{equation}\label{t6-50.3}
  \frac{df}{dt}=\frac{2}{\phi_0}\sqrt{\frac{f-1}{f}},\qquad f(0)=1
\end{equation}
and it can be defined in implicit form by the equation
\begin{equation}\label{h13}
  t(f)=\frac{\phi_0}{2}\left[\sqrt{f(f-1)}+\ln(\sqrt{f-1}+\sqrt{f})\right].
\end{equation}

Let the dark soliton be formed at the initial moment of time $t=0$ at the point $x=0$ with
the velocity $V_0>0$. Then its path can be found by solving the equation (\ref{t6-47.14})
which in our case reads
\begin{equation}\label{t6-50.4}
  \frac{dx}{dt}-\frac12g(t)x=V_0,
\end{equation}
with the initial condition $x(0)=0$. This linear differential equation can be easily solved
and with the use of the expressions for $g(t)$ and $df/dt$ the solution can be cast to the form
\begin{equation}\label{t6-50.5}
  x(f)=V_0\phi_0\sqrt{f(f-1)}.
\end{equation}
This formula together with Eq.~(\ref{h13}) define the soliton's path $x=x(t)$ in parametric form.
The soliton's velocity is to be found from Eq.~(\ref{t6-50.4}),
\begin{equation}\label{t6-50.6}
  \dot{x}(f)=V_0\left(2-\frac1f\right).
\end{equation}
Typical condensate profiles at different moments of time $t$ are shown in Fig.~\ref{fig1}
for $\mu=1$, $\omega_0=0.1$ and $V_0=0.5$.
Plots of $x(t)$ and $V=\dot{x}(t)$ are shown in Fig.~\ref{fig2}. In the limit $t\to\infty$,
when $f\to\infty$, the velocity tends to its limiting value $\dot{x}(\infty)=2V_0$ equal to
the value of the flow velocity $u= x/t=2V_0$ along the limiting path $x(t)= 2V_0t$.
Consequently, at asymptotically large times the soliton becomes black and it is transferred by
the flow.

\section{Dark soliton's motion along a hydraulic flow of BEC through a penetrable barrier}\label{sec3}

\begin{figure}[t]
	\begin{center}
		\includegraphics[width=8cm]{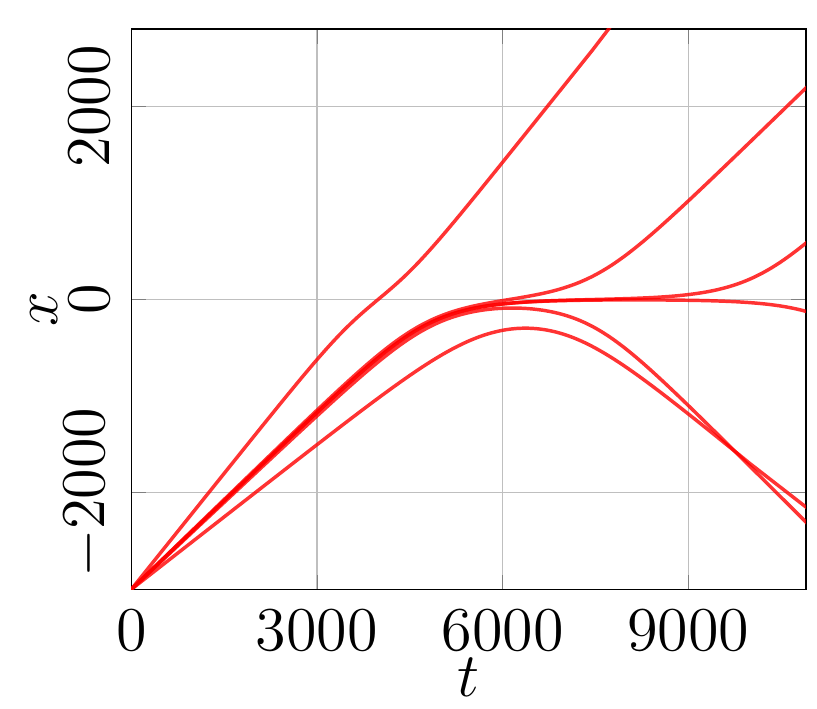}
		\caption{Soliton's paths for fixed values of the parameters $\rho_0$, $u_0$ of the
			flow at infinity $x\to\pm\infty$, the fixed potential barrier shape and different
			values of the soliton's initial velocity $V_0$. At some critical value of the velocity $V_0$ the paths passing
			through the barrier change to the paths reflected from it. The plot corresponds to the values
			$\rho_0=1.0$, $u_0=0.1$, $U_m=0.4$, and $\sigma=300$ (see Eq.~(\ref{eq17})).}
		\label{fig3}
	\end{center}
\end{figure}

\begin{figure*}[t]
	\begin{center}
		\includegraphics[width=\linewidth]{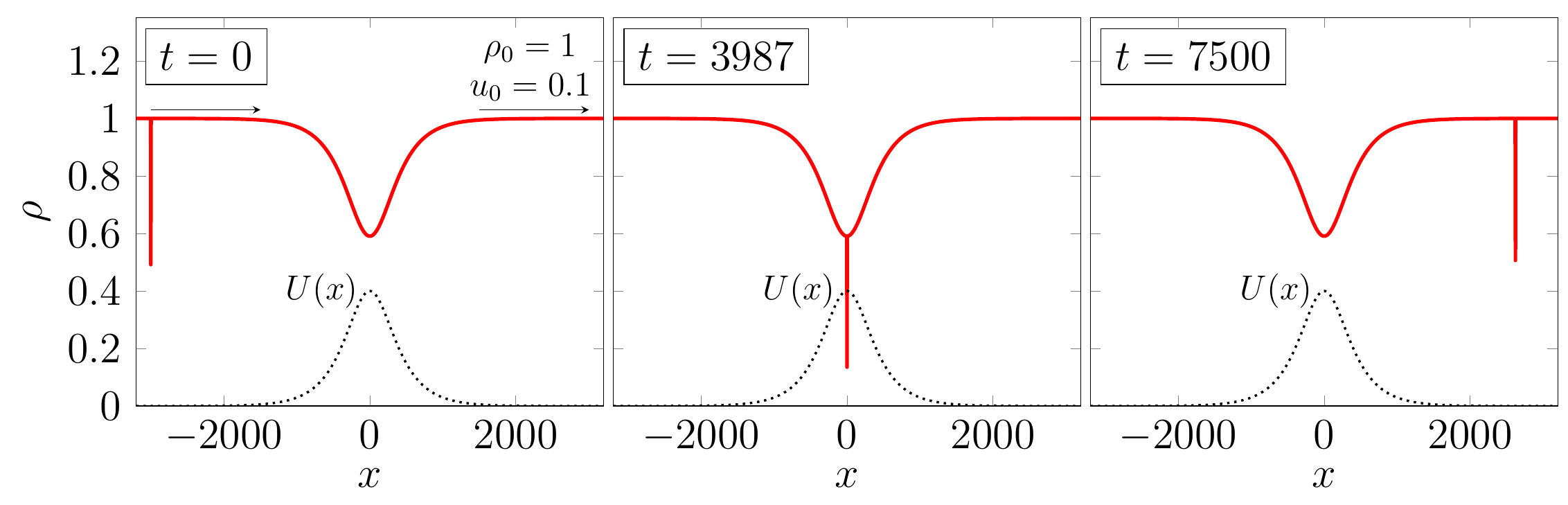}
		\caption{Downstream propagation of a dark soliton through the non-uniform
			flowing condensate.
			Red curve shows the density $\rho(x)$ and the dotted line is the potential $U(x)$.
			The background density at infinity is taken equal to $\rho_0=1$ and flow velocity is $u_0=0.1$,
			the initial soliton's velocity is $V_s=0.8$. The potential has a shape~(\ref{eq17})
			with the maximal value $U_m=0.4$
			and the width $\sigma=300$. The arrows show direction of the background flow.}
		\label{fig4}
	\end{center}
\end{figure*}

\begin{figure}[t]
	\begin{center}
		\includegraphics[width=0.7\linewidth]{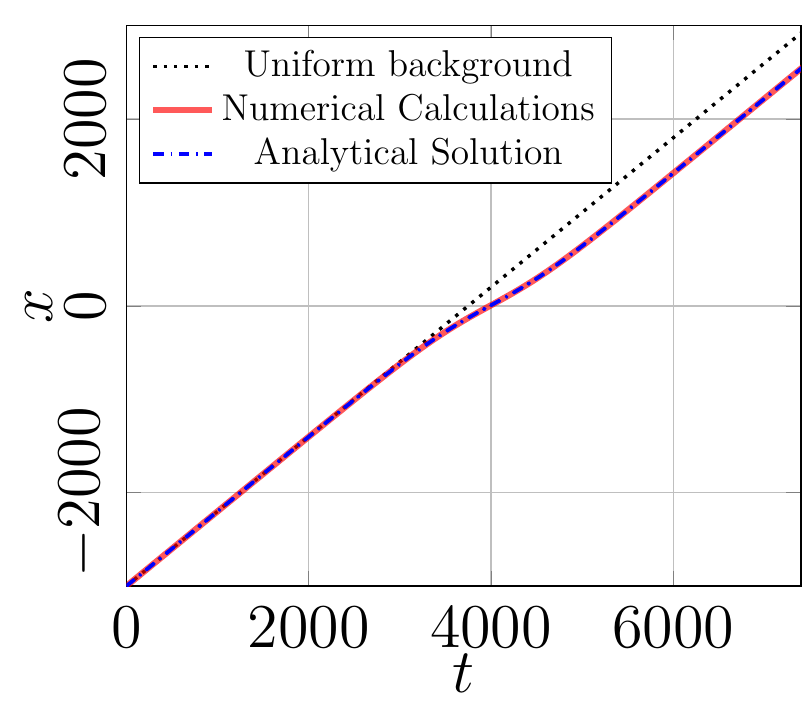}
		\caption{The dependence of the soliton position on time for the dynamics
			shown in Fig.~\ref{fig4}. The dash-dotted blue line corresponds to the solution of Eq.~(\ref{t6-47.12b})
			and the red line to the numerical solution of the GP
			equation (\ref{eq1}). The dotted line shows the soliton's path through a uniform
			condensate when $U_m=0$.}
		\label{fig5}
	\end{center}
\end{figure}

Now we shall consider a stationary flow of BEC past a wide barrier whose action is described by
the potential $U(x)$.  The large scale dependence of the density $\rho$ and the flow velocity $u$
is determined by the stationary version of Eqs.~(\ref{t6-47.11}),
\begin{equation}\label{eq5}
	(\rho u)_x=0,\quad uu_x+\rho_x+U_x=0,
\end{equation}
which are readily integrated to give
\begin{equation}
	\rho u=\rho_0 u_0,\quad \frac12 u^2+\rho+U=\frac12 u_0^2+\rho_0,
\end{equation}
where $\rho_0$ and $u_0$ are the density and the flow velocity of the condensate at $x\to\pm\infty$.
We imply here that the potential is localized around the point $x=0$ and its distance of action
$\sigma$ is much greater than the healing length which is equal to unity in our dimensionless notation.
Elimination of $\rho=\rho_0u_0/u$ yields the equation
\begin{equation}\label{eq6}
	\frac{1}{2}\left[u_0^2-u^2(x)\right]+\rho_0\left(1-\frac{u_0}{u(x)}\right)=U(x),
\end{equation}
which defines in implicit form the function $u=u(x)$ for the space dependence of the background
flow velocity. When it is found, the distribution of the background density is given by the formula
\begin{equation}\label{eq6b}
  \rho(x)=\frac{\rho_0u_0}{u(x)}.
\end{equation}
Equation (\ref{eq6}) is cubic and its solution should be chosen in such a way that it satisfies
the boundary condition $u(x)\to u_0$ at both infinities $x\to\pm\infty$. As was noticed in
Refs.~\cite{hakim-97,legk-09}, this imposes a very important condition on possible values of $u_0$.
Indeed, at the point where the potential $U(x)$ reaches its maximal value
$U_m =\max\{ U(x):-\infty<x<\infty\}$,
the function in the left-hand side of Eq.~(\ref{eq6}) has the maximum equal to
$\frac12u_0^2-\frac32(\rho_0u_0)^{2/3}+\rho_0$ at $u(x)=u_m=(u_0\rho_0)^{1/3}$.
Consequently, the solution $u(x)$ for all
values of $x$, $-\infty<x<\infty$, exists if only
\begin{equation}\label{eq7}
	U_m\le\frac12 u_0^2-\frac32\left(\rho_0u_0\right)^{2/3}+\rho_0.
\end{equation}
This inequality becomes equality for $u_0$ equal to two roots $u_{\pm}$ of the equation
\begin{equation}\label{}
	\frac12u_0^2-\frac32(\rho_0u_0)^{2/3}+\rho_0=U_m.
\end{equation}
Then it is easy to find that the smooth solution $u=u(x)$ exists for all $x$ in
the subcritical $u_0<u_-$ and supercritical $u_0>u_+$ regimes of the flow past an
obstacle described by the potential $U(x)$. We assume that $u_0>0$, so the root
$u_-$ should be put equal to zero for $U_m\geq\rho_0$. If $U_m\ll\rho_0$, then $u_0$ must be
close to the asymptotic value of the sound velocity $\sqrt{\rho_0}$, so one can easily
find that up to the second order in the small parameter $(U_m/\rho_0)^{1/2}$ we have
\begin{equation}\label{eq8}
	u_\pm\approx \sqrt{\rho_0}\left(1\pm\sqrt{\frac{3U_m}{2\rho_0}}+\frac{U_m}{12\rho_0}\right).
\end{equation}
In the transcritical regime $u_-<u_0<u_+$, the condition (\ref{eq7}) does not
hold. In this case, the flow past a wide barrier leads to generation of
dispersive shock waves in this finite interval of the flow velocities \cite{hakim-97,legk-09}.
We are only interested in the situations when dispersive shock waves are not formed
and the soliton passes through a stationary fairly smooth profile of the condensate.

In our concrete examples we shall use the potential
\begin{equation}\label{eq17}
	U(x)=\frac{U_m}{\cosh(x/\sigma)},
\end{equation}
where $\sigma=300$ and $U_m=0.4$. We take the initial coordinate of the dark soliton $x_0$
upstream the flow far away from the barrier and launch the dark soliton with the initial
velocity $V_0$. Soliton's motion obeys the Newton equation (\ref{t6-47.12b}) which can be
easily solved numerically with the initial conditions $x(0)=x_0$, $\dot{x}(0)=V_0$.

\begin{figure*}[t]
	\begin{center}
		\includegraphics[width=\linewidth]{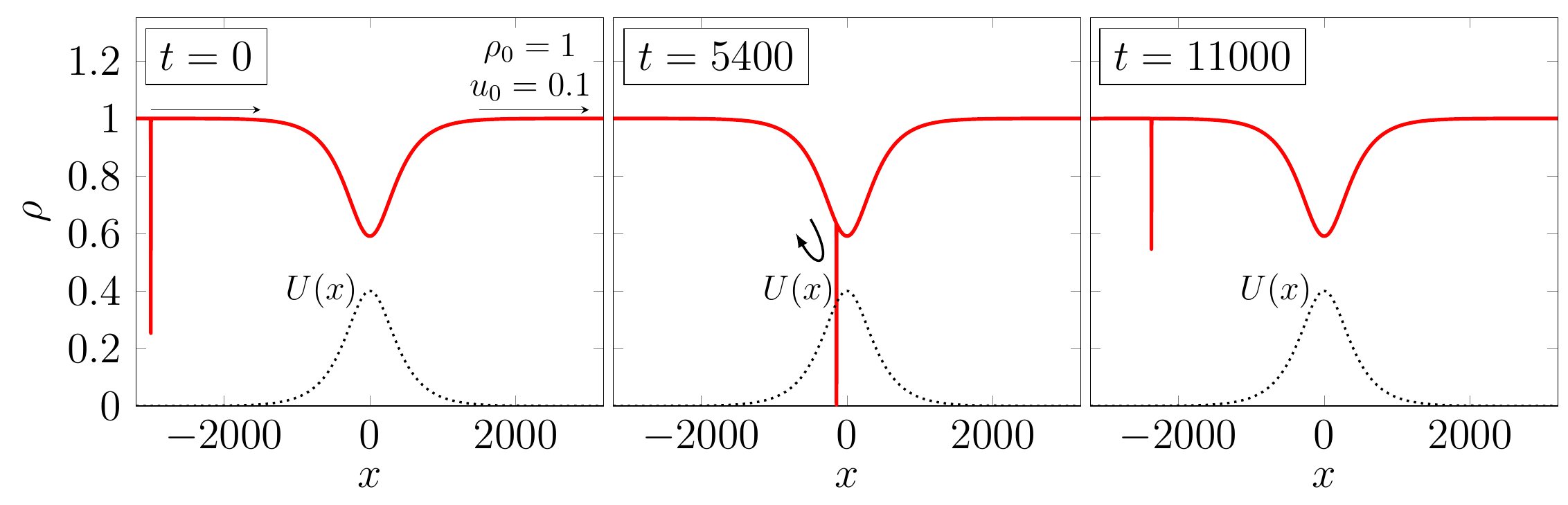}
		\caption{Reflection of a dark soliton from the potential barrier. Red curve shows the density
$\rho(x)$, and the dotted line depicts the potential $U(x)$. Background density is equal to $\rho_0=1$
and flow velocity to $u_0=0.1$, initial soliton velocity is $V_s=0.6$. The potential has the form
(\ref{eq17}) with $U_m=0.4$ and $\sigma=300$.}
		\label{fig6}
	\end{center}
\end{figure*}

\begin{figure}[t]
	\begin{center}
		\includegraphics[width=0.7\linewidth]{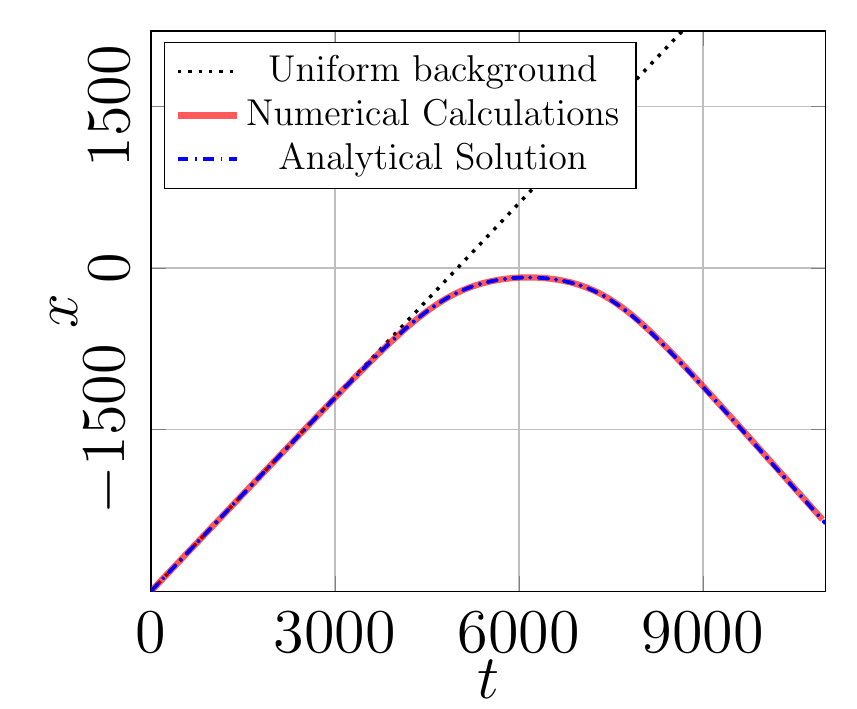}
		\caption{The path of the soliton for its propagation for the dynamics
			shown in Fig.~\ref{fig6}.}
		\label{fig7}
	\end{center}
\end{figure}

\begin{figure}[t]
	\begin{center}
		\includegraphics[width=0.7\linewidth]{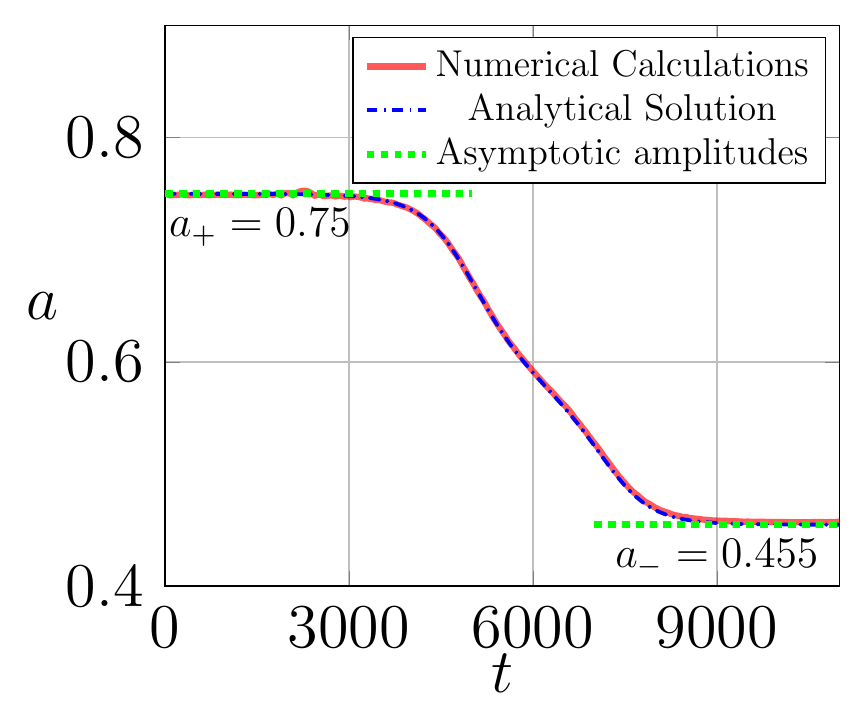}
		\caption{Soliton amplitude for the dynamics
			shown in Fig.~\ref{fig6}. The green dotted lines correspond to the asymptotic amplitudes
of the soliton, which are in agreement with Eq.~(\ref{eq6-24}).}
		\label{fig8}
	\end{center}
\end{figure}

First of all, we notice that there exist two types of solitons trajectories: (i) trajectories which
pass through the potential region from one infinity to the other one, and (ii) trajectories with reflection
of solitons from the potential region. This is illustrated in Fig.~\ref{fig3}. We fix the parameters
$\rho_0$, $u_0$ of the flow at infinity $x\to\pm\infty$ and calculate solitons paths for different
values of the soliton's initial velocity $V_0$ in the range from $0.5$ to $0.8$.
At some critical value of the soliton's velocity $V_0\approx0.61$ the paths switch from those passing through
the barrier to the paths reflected from it.
At the critical value $a_0 = a_{cr}$, which separates these two regimes, the trajectory reaches the
point $x = 0$ of the potential maximum with zero velocity, that is $\dot{x}(x = 0) = 0$ at the point
with $u(0) = u_m, \rho(0) = \rho_m$. Substitution of these values of into Eq.~(\ref{h2}) gives the
soliton's critical energy
\begin{equation}\label{crit-energy}
\begin{split}
  \eps_{cr}=&\frac2{3\rho_m^3}(2\rho_m^3+\rho_0^2u_0^2)\sqrt{\rho_m^3-\rho_0^2u_0^2}\\
  &+2u_0\rho_0\left(\frac{\pi}2+\arcsin\frac{\rho_0u_0}{\rho_m^{3/2}}\right).
  \end{split}
\end{equation}
Equating this soliton's energy at the turning point to its initial energy, we arrive at the equation
\begin{equation}\label{eq107.4}
\begin{split}
  \frac43a_{cr}^{3/2}-2u_0\sqrt{a_{cr}(\rho_0-a_{cr})}+2u_0\rho_0\arcsin\sqrt{\frac{a_{cr}}{\rho_0}}=\eps_{cr}.
  \end{split}
\end{equation}
If we define the variables
\begin{equation}\label{eq5-20}
  A=\frac{a_{cr}}{\rho_0},\qquad M=\frac{u_0}{\sqrt{\rho_0}},
\end{equation}
then we can write this equation in the form
\begin{equation}\label{eq5-21}
  M(A)=\frac{4A^{3/2}-3\eps_{cr}/\rho_0^{3/2}}{6\left(\sqrt{A(1-A)}-\arcsin\sqrt{A}\right)}.
\end{equation}
It defines in implicit form the dependence of the non-dimensional critical amplitude on the Mach number
of the flow far from the obstacle. In particular, for the values $\rho_0=1.0$, $u_0=0.1$, $U_m=0.4$,
we obtain $\rho_{min}=0.591$ and Eq.~(\ref{eq5-21}) yields $A_{cr}=0.738$ ($V_0=0.612$) in reasonable agreement
with the value obtained in numerical solutions of the GP equation.

If we turn to the trajectories with reflection of solitons from the obstacle, then we see that the
initial and final amplitudes of such a soliton are different due to change of sign of the velocity:
\begin{equation}\label{eq6-22}
  a_+=\rho_0-(V_+-u_0)^2,\quad a_-=\rho_0-(V_--u_0)^2,
\end{equation}
where both velocities are defined far enough from the obstacle. Let the soliton start its downstream
motion at $x=-\infty$ with the velocity $V_+>0$, so its initial energy is given by the expression
\begin{equation}\nonumber
  \eps=\frac43a_+^{3/2}-2u_0\sqrt{a_+(\rho_0-a_+)}+2u_0\rho_0\arcsin\sqrt{\frac{a_+}{\rho_0}},
\end{equation}
and it must be equal to its energy after reflection when it moves upstream with the velocity $V_-<0$
so that its energy is given by
\begin{equation}\nonumber
	\begin{split}
  \eps=\frac43a_-^{3/2}+&2u_0\sqrt{a_-(\rho_0-a_-)} \\
  +&2u_0\rho_0\left(\pi-\arcsin\sqrt{\frac{a_-}{\rho_0}}\right).
  \end{split}
\end{equation}
Equating these two expressions and introducing again the variables $A_+=a_+/\rho_0$, $A_-=a_-/\rho_0$,
$M=u_0/\sqrt{\rho_0}$, we get the equation
\begin{equation}\label{eq6-24}
\begin{split}
  \frac23A_-^{3/2}&+M\sqrt{A_-(1-A_-)}+M\left(\pi-\arcsin\sqrt{A_-}\right)\\
  &=\frac23A_+^{3/2}-M\sqrt{A_+(1-A_+)}+M\arcsin\sqrt{A_+},
  \end{split}
\end{equation}
which determines the final amplitude of the soliton in terms of its initial amplitude and the
Mach number of the flow.

We compared our analytical findings with numerical solutions of the GP equation.
An example of soliton's propagation for a subcritical ($u_0<u_-$) background flow velocity is shown in
Fig.~\ref{fig4}. The soliton moves downstream the condensate flow.
Fig.~\ref{fig5} shows the soliton's trajectory calculated
according to Eq.~(\ref{t6-47.12b}) and by means of numerical solution of the GP
equation (\ref{eq1}). As one can see, the approximate analytical theory (dash-dotted blue line) agrees perfectly well
with the exact numerical solution (red line). Similar agreement between two approaches was found
also for supercritical flow of the background BEC and for the upstream initial motion of solitons.

Another case, when the soliton is reflected from the barrier, is illustrated in Fig.~\ref{fig6}.
The solution of Eq.~(\ref{t6-47.12b}) agrees perfectly well with
the exact numerical solution of the GP equation, as is shown in Fig.~\ref{fig7}.
One can clearly see that the soliton's amplitude changes after reflection.
Fig.~\ref{fig8} shows the dependence of the amplitude on $t$. As one can see, the numerical
amplitude of the soliton after the collision coincides with the analytical one obtained from Eq.~(\ref{eq6-24}).


\section{Conclusion}\label{sec5}

The motion of dark solitons is an interesting problem of nonlinear physics because such a motion
cannot be completely separated from the motion of the background medium. Most spectacularly this
phenomenon is demonstrated by the difference in the motion of the center of mass of BEC confined
in a trap and the motion of the soliton itself described by its mean coordinate.
Consequently, the soliton's motion
is accompanied by some counterflow in the background condensate which leads to a proper
redistribution of the density. In 1D geometry such a counterflow is necessary for topological
reasons: the flow velocity of the condensate is a gradient of the phase of the condensate's
wave function, so formation of a dark soliton leads to the jump of phase across the
soliton and, to keep the wave function single-valued, this jump must be compensated by the flow
outside the soliton \cite{shevchenko-88}. We show that the contribution of the counterflow into the
momentum and the energy of a dark soliton is crucially important for correct description of
its motion as a localized quasi-particle by the methods of Hamiltonian mechanics. In this way
we have reproduced the results obtained earlier by the methods of perturbation theory \cite{ba-2000}
and in framework of the Whitham theory \cite{she-18} (in the last case only for situations without
external potential). The advantage of the Hamiltonian approach is that it provides immediately
the quite non-trivial energy conservation law for the soliton's motion in stationary background
flows and it would be difficult to find this law by other methods. We have illustrated our approach by
several examples and confirmed the analytical results by their comparison with numerical solutions
of the GP equation.

We believe that the suggested here Hamiltonian approach can be applied to other interesting
situations, such as, for example, interaction of solitons with non-convex flows \cite{seh-21}
or ``magnetic'' solitons \cite{qps-16,pit-16} in BEC, when the main contribution into the soliton's
canonical momentum is made by the counterflow.


\begin{acknowledgments}

We thank D.~V.~Shaykin for useful discussions. This work was partially supported by the Foundation for
the Advancement of Theoretical Physics and Mathematics ``BASIS.''

\end{acknowledgments}


\begin{thebibliography}{99}

\bibitem{zk-08} N. J. Zabusky, M. D. Kruskal,  Phys. Rev. Lett. {\bf 15,} 240 (1965).

\bibitem{km-77} V. I. Karpman, E. M. Maslov, 
Sov. Phys. JETP, {\bf 46,} 281 (1977).

\bibitem{km-78} V. I. Karpman, E. M. Maslov,
Sov. Phys. JETP, {\bf 48,} 252 (1978).

\bibitem{kn-78} D. J. Kaup and A. C. Newell,
Proc. Roy. Soc. London,  A {\bf 361,} 413 (1978).

\bibitem{kosevich-90} A.~M.~Kosevich, Physica D {\bf 41,} 253 (1990).

\bibitem{ba-2000} Th.~Busch, J.~R.~Anglin, 
Phys. Rev. Lett. {\bf 84,} 2298-2301 (2000).

\bibitem{shevchenko-88} S.~I.~Shevchenko, Sov. J. Low Temp. Phys. {\bf 14,} 553 (1988).

\bibitem{ps-2003} L.~P.~Pitaevskii, S.~Stringari, {\it Bose-Einstain Condensation,} (Clarendon, Oxfors, 2003).

\bibitem{ky-94} Yu.~S.~Kivshar, X.~Yang, Phys. Rev. E {\bf 49,} 1657 (1994).

\bibitem{kk-95} Yu.~S.~Kivshar, W. Kr\'{o}likowski, 
Optics Commun., {\bf 114,} 353-362 (1995).

\bibitem{kp-04} V.~V.~Konotop, L.~P.~Pitaevskii, 
Phys. Rev. Lett. {\bf 93,} 240403 (2004).

\bibitem{kk-10} A.~M.~Kamchatnov, S.~V.~Korneev, 
Phys. Lett. A, {\bf 374,} 4625 (2010).

\bibitem{she-18} P.~Sprenger, M.~A.~Hoefer, G.~A.~El, 
Phys. Rev. E {\bf 97,} 032218 (2018).

\bibitem{talanov-65}  V. I. Talanov, JETP Lett., {\bf 2,} 138 (1965).

\bibitem{bkk-03}  V. A. Brazhnyi, A. M. Kamchatnov,  V. V. Konotop, Phys. Rev. A {\bf 68,} 035603 (2003).

\bibitem{hakim-97} V.~Hakim, Phys. Rev. E {\bf 55,} 2835 (1997).

\bibitem{legk-09} A.~M.~Leszczyszyn, G.~A.~El, Yu.~G.~Gladush, A.~M.~Kamchatnov,
Phys. Rev. A {\bf 79,} 063607 (2009).

\bibitem{kamch-2000} A.~M.~Kamchatnov, {\it Nonlinear Periodic Waves and Their Modulations---An Introductory Course}, (World Scientific, Singapore, 2000).

\bibitem{becker-08} Ch.~Becker, S.~Stellmer, P.~Soltan-Panahi, S.~D\"{o}rscher, M.~Baumert,
E.-M.~Richter, J.~Kronj\"{a}ger, K.~Bongs, K.~Sengstock, 
Nature Phys., {\bf 4,} 496-501 (2008).

\bibitem{weller-08} A.~Weller, J.~P.~Ronzheimer, C.~Gross, J.~Esteve, M.~K.~Oberthaler, D.~J.~Frantzeskakis,
G.~Theocharis, P.~G.~Kevrekidis, 
Phys. Rev. Lett., {\bf 101,} 130401 (2008).

\bibitem{seh-21} K. van der Sande, G.~A.~El, M.~A.~Hoefer, 
J. Fluid Mech., {\bf 928,} A21 (2021).

\bibitem{qps-16} C. Qu, L. P. Pitaevskii, and S. Stringari, 
Phys. Rev. Lett. {\bf 116,} 160402 (2016).

\bibitem{pit-16} L. P. Pitaevskii, Phys.-Uspekhi, {\bf 59,} 1028 (2016).





\end{thebibliography}
\end{document}